\documentclass[twocolumn]{aastex63}
\usepackage{mathtools}
\usepackage{dsfont}
\DeclareMathOperator*{\argmax}{arg\,max}

\received{}
\revised{}
\accepted{}

\graphicspath{{./}{figures/}}
\defcitealias{moon21}{Paper~I}

\shorttitle{Episodicity and Asymmetry of Nuclear Rings}
\shortauthors{Moon et al.}

\begin{document}

\title{Effects of Varying Mass Inflows on Star Formation in Nuclear Rings of Barred Galaxies}

\author[0000-0002-6302-0485]{Sanghyuk Moon}
\affiliation{Department of Physics \& Astronomy, Seoul National University, 1 Gwanak-ro, Gwanak-gu, Seoul 08826, Republic of Korea}
\affiliation{SNU Astronomy Research Center, Seoul National University, 1 Gwanak-ro, Gwanak-gu, Seoul 08826, Republic of Korea}
\author[0000-0003-4625-229X]{Woong-Tae Kim}
\affiliation{Department of Physics \& Astronomy, Seoul National University, 1 Gwanak-ro, Gwanak-gu, Seoul 08826, Republic of Korea}
\affiliation{SNU Astronomy Research Center, Seoul National University, 1 Gwanak-ro, Gwanak-gu, Seoul 08826, Republic of Korea}
\author[0000-0003-2896-3725]{Chang-Goo Kim}
\affiliation{Department of Astrophysical Sciences, Princeton University,
  Princeton, NJ 08544, USA}
\author[0000-0002-0509-9113]{Eve C.\ Ostriker}
\affiliation{Department of Astrophysical Sciences, Princeton University,
  Princeton, NJ 08544, USA}

\email{s.moon@snu.ac.kr, wkim@astro.snu.ac.kr, cgkim@astro.princeton.edu, eco@astro.princeton.edu}

\begin{abstract}
Observations indicate that the star formation rate (SFR) of nuclear rings varies considerably with time and is sometimes asymmetric rather than being uniform across a ring.
To understand what controls temporal and spatial distributions of ring star formation, we run semi-global, hydrodynamic simulations of nuclear rings subject to time-varying and/or asymmetric mass inflow rates.
These controlled variations in the inflow lead to variations in the star formation, while the ring orbital period ($18\,{\rm Myr}$) and radius ($600\,{\rm pc}$) remain approximately constant.
We find that both the mass inflow rate and supernova feedback affect the ring SFR. An oscillating inflow rate with period $\Delta \tau_\text{in}$ and amplitude 20 causes large-amplitude (a factor of $\gtrsim 5$), quasi-periodic variations of the SFR, when $\Delta \tau_\text{in} \gtrsim 50\,{\rm Myr}$.
We find that the time-varying ISM weight and midplane pressure track each other closely, establishing an instantaneous vertical equilibrium. The measured time-varying depletion time is consistent with the prediction from self-regulation theory provided the time delay between star formation and supernova feedback is taken into account.
The supernova feedback is responsible only for small-amplitude (a factor of $\sim 2$) fluctuations of the SFR with a timescale $\lesssim 40\,{\rm Myr}$.
Asymmetry in the inflow rate does not necessarily lead to asymmetric star formation in nuclear rings.
Only when the inflow rate from one dust lane is suddenly increased by a large factor, the rings undergo a transient period of lopsided star formation.
\end{abstract}

\section{Introduction}\label{s:intro}

Nuclear rings at the centers of barred spiral galaxies are conspicuous in ultraviolet or H$\alpha$, indicative of on-going massive star formation activity \citep{maoz96,benedict02}.
The ring star formation rate (SFR) inferred from H$\alpha$ luminosity is in the range of $0.1$--$10\,{M_\odot\,{\rm yr}^{-1}}$ \citep{mazzuca08,ma18}, which is high enough to produce pseudobulges with masses of $10^8$--$10^{10}\,M_\odot$ within $1\,{\rm Gyr}$, provided that the ring is continuously supplied with fresh gas \citep{kormendy04}.
Both observations and simulations indicate that in barred galaxies, gas falls into the nuclear ring through a pair of dust lanes located on the leading side of a bar \citep[e.g.,][]{athanassoula92,benedict96,regan97,regan99,laine99,schinnerer02,wtkim12a,sormani15,sormani19,shimizu19}.
The mass inflow rates through the dust lanes inferred by kinematic observations are $1$--$7\,{M_\odot\,{\rm yr}^{-1}}$ \citep{benedict96,regan97,laine99,sormani19,shimizu19}, similar to the observed range of the ring SFR.\footnote{\citet{hatchfield21} showed
that only $\sim 30\%$ of the inflowing gas along the dust lanes directly land on the ring, while the rest overshoots and accrete at later passages \citep[see also,][]{regan97}, making the true accretion rate smaller by a factor of about 3 than the inferred rate.}

Despite numerous studies mentioned above, it is still uncertain what controls star formation in nuclear rings. 
On the one hand, \citet{kruijssen14} \citep[see also,][]{loose82,krugel93,elmegreen94,krumholz17,torrey17} proposed that star formation in nuclear rings is strongly episodic.
In this scenario, inflowing gas gradually piles up in a ring.
When the ring becomes massive enough, it undergoes gravitational instability and this leads to an intense burst of star formation.
The related strong stellar feedback quenches further star formation, rendering the ring quiescent until it once again becomes massive enough to become unstable.
This cycle repeats, and the resulting ring SFR is episodic with periods of $20$--$40\,{\rm Myr}$ \citep{krumholz17}.
On the other hand, global simulations of \citet{wyseo13,wyseo14} and \citet{wyseo19} found that the SFR history in the nuclear ring is very similar to the history of the mass inflow rates through the bar, suggesting that the ring SFR is primarily determined by the inflow rate.

In global simulations of nuclear rings \citep[e.g.,][]{wyseo19,armillotta19,sormani20,tress20}, the inflow rate is naturally time variable  and feedback is always active following star formation,
so that it is difficult to discern what dominates in shaping the star formation history of nuclear rings. 
To separate the effects of feedback from those of the mass inflow, \citet[hereafter Paper I]{moon21} designed semi-global models that focus on nuclear rings and nearby regions, handling the bar-driven mass inflows along dust lanes by gas streams through two nozzles located at the domain boundaries.
A key advantage of this semi-global framework is the ability to adjust the mass inflow rate as a free parameter.   
Based on simulations with the inflow rates kept constant in time,  \citetalias{moon21} found that the ring SFR is tightly correlated with the inflow rate and that the midplane pressure powered by supernova (SN) feedback balances the weight of the overlying gas.
In these simulations, SN feedback never destroys the rings completely and induces only modest (within a factor of $\sim2$) temporal fluctuations of the SFR.

While the models considered in \citetalias{moon21} are informative in understanding ring star formation for a constant inflow rate, the mass inflows in real barred galaxies result from dynamical interactions of gas with a stellar bar and are thus likely time variable.
For example, \citet{sormani19} measured the density and velocity of the gas associated with the dust lanes of the Galactic bar and predicted that the mass inflow rate to the central molecular zone (CMZ), the nuclear ring in the Milky Way, should display a factor of few variations in the future $\sim 12\,{\rm Myr}$.
Global numerical simulations \citep[e.g.,][]{wyseo19,armillotta19,tress20} also found that the mass inflow rate varies by more than an order of magnitude over timescales of a few tens to hundreds of ${\rm Myr}$, potentially responsible for the observed episodic star formation in nuclear rings \citep{allard06,sarzi07,gadotti19,prieto19}. To understand the history of the ring SFR, it is therefore desirable to run models with time-dependent mass inflow rates.

Another important issue regards lopsided star formation in nuclear rings.
Although a majority of nuclear rings appear more or less symmetric in the distributions of star-forming regions, some rings show clear asymmetric star formation around their circumference \citep{comeron10,ma18}.
Notable examples of asymmetric star formation include the CMZ \citep{bally88,henshaw16} and the nuclear ring of M83 \citep{harris01,callanan21}, in which star formation is not uniformly distributed but concentrated roughly in a quarter-to-half portion of the ring.
Possible causes for such asymmetry include a recent minor merger potentially responsible for an offset between the photometric and kinematic nucleus \citep[e.g.,][]{sakamoto04,knapen10} and asymmetric mass inflow along the two dust lanes, the latter of which can readily be checked by direct numerical simulations using the framework presented in \citetalias{moon21}.

In this paper, we extend \citetalias{moon21} to investigate how time-varying and/or asymmetric inflow rates affect temporal and spatial variations of the ring SFR.
We consider two series of models.
In the first series, the inflow rate from two nozzles is symmetric but oscillates in time with period $\Delta \tau_\text{in}$.
By running models with the same time-averaged rate but differing $\Delta \tau_\text{in}$, we study the relationship between $\Delta \tau_\text{in}$ and the time variations of the SFR.
In the second series, the inflow rate of the two nozzles is forced to be asymmetric, either from the beginning or suddenly at a specified time.
By comparing three representative cases of asymmetric inflows, we find conditions for lopsided star formation in nuclear rings.

The remainder of this paper is organized as follows.
In Section \ref{s:method}, we briefly describe our semi-global models and numerical methods to evolve the models subject to radiative heating and cooling, star formation, and SN feedback.
In Section \ref{s:results}, we present the numerical results for the temporal and spatial distributions of star formation in nuclear rings.
We summarize and discuss our results in Section \ref{s:discussion}.

\section{Methods}\label{s:method}

In  \citetalias{moon21}, the mass inflow rates through two nozzles were kept symmetric and constant with time. 
Here we study ring star formation when the mass inflow rate varies with time or becomes asymmetric.
In this section, we briefly summarize the equations we solve  and the treatment of star formation and feedback: we refer the reader to \citetalias{moon21} for a complete description of our semi-global models of nuclear rings.

\subsection{Equations}\label{s:equations}

Our computational domain is a Cartesian cube with side $L=2048\,{\rm pc}$ surrounding a nuclear ring.
We discretize the domain uniformly with $512^3$ cells with grid spacing of $\Delta x = 4\,{\rm pc}$.
The domain rotates at an angular frequency ${\bf \Omega}_p=36\,{\rm km\,s^{-1}\,kpc^{-1}}\,\hat{\bf z}$, representing the pattern speed of a bar.
We use the {\tt Athena} code \citep{stone08} to integrate the equations of hydrodynamics in the rotating frame, coupled with self-gravity, heating and cooling, star formation, and SN feedback. 
The governing equations we solve read 
\begin{equation}\label{eq:cont}
    \frac{\partial\rho}{\partial t} + \boldsymbol\nabla\cdot\left(\rho {\bf v}\right) = 0,
\end{equation}
\begin{equation}
\begin{split}
    \frac{\partial(\rho {\bf v})}{\partial t} + \boldsymbol\nabla\cdot\left(\rho {\bf v}{\bf v} + P\mathds{I}\right) = - 2\rho{\bf \Omega}_p\times{\bf v}-\rho\boldsymbol\nabla\Phi_{\rm tot},
\end{split}
\end{equation}
\begin{equation}\label{eq:energy}
\begin{split}
    \frac{\partial}{\partial t}\left(\frac{1}{2}\rho v^2 + \frac{P}{\gamma-1}\right) + \boldsymbol\nabla\cdot\left[\left(\frac{1}{2}\rho v^2 + \frac{\gamma P}{\gamma - 1}\right){\bf v}\right]\\= -\rho{\bf v}\cdot\boldsymbol\nabla\Phi_{\rm tot} - n_{\rm H}^2\Lambda + n_{\rm H}\Gamma_{\rm PE} + n_{\rm H}\Gamma_{\rm CR},
\end{split}
\end{equation}
where $\rho$ is the gas density, ${\bf v}$ is the gas velocity in the rotating frame, $P$ is the gas pressure, $\mathds{I}$ is the identity matrix, $n_{\rm H}=\rho/(1.4271 m_{\rm H})$ is the hydrogen number density assuming the solar abundances, $n_{\rm H}^2\Lambda$ is the volumetric cooling rate, $n_{\rm H}\Gamma_{\rm PE}$ is the photoelectric heating rate, and $n_{\rm H}\Gamma_{\rm CR}$ is the heating rate by cosmic ray ionization.
We assume that the cooling function $\Lambda$ depends only on the gas temperature $T$, and adopt the forms suggested by \citet{koyama02} for $T<10^{4.2}\,{\rm K}$ and \citet{sutherland93} for $T>10^{4.2}\,{\rm K}$.

The total gravitational potential $\Phi_{\rm tot}=\Phi_{\rm cen} + \Phi_{\rm ext} + \Phi_{\rm self}$ consists of the centrifugal potential $\Phi_{\rm cen}=-\frac{1}{2}\Omega_p^2(x^2+y^2)$, the external gravitational potential $\Phi_{\rm ext}$ giving rise to the background rotation curve, and the self-gravitational potential $\Phi_{\rm self}$ that is related to $\rho$ and the newly-formed star particle density $\rho_{\rm sp}$ via
\begin{equation}\label{eq:Poisson}
    \boldsymbol\nabla^2\Phi_\text{self} = 4\pi G(\rho + \rho_{\rm sp}).
\end{equation}
The background potential $\Phi_{\rm ext}$ arises from the central supermassive black hole modeled by the Plummer sphere
\begin{equation}
    \rho_{\rm BH} = \frac{3M_{\rm BH}}{4\pi r_{\rm BH}^3}\left(1 + \frac{r^2}{r_{\rm BH}^2}\right)^{-5/2}
\end{equation}
with mass $M_{\rm BH} = 1.4\times 10^8\,M_\odot$ and the softening radius $r_{\rm BH} = 20\,{\rm pc}$, and a spherical stellar bulge modeled by the modified Hubble profile
\begin{equation}\label{eq:bulge}
    \rho_b(r) = \frac{\rho_{b0}}{(1+r^2/r_b^2)^{3/2}}
\end{equation}
with the central density $\rho_{b0} = 50\,M_\odot\,{\rm pc}^{-3}$ and the scale radius $r_b = 250\,{\rm pc}$.
The total stellar mass enclosed within the nuclear ring is $M_b \equiv \int_0^{R_{\rm ring}}4\pi r^2\rho_b(r)dr = 6.7\times 10^9\,M_\odot$, where $R_{\rm ring}=600\,{\rm pc}$ is the ring radius\footnote{In our semi-global models, the ring radius is set by the velocity (or, more precisely, the specific angular momentum) of the inflows, as indicated by Equation (12) of \citetalias{moon21}.}.
The rotation curve resulting from $\Phi_{\rm ext}$ is similar to the observations of a prototypical nuclear-ring galaxy NGC 1097 (\citealt{onishi15}; see also Figure 1 of \citetalias{moon21}). With these parameters, the orbital period (in rotating frame) at $R_{\rm ring}$ is $t_{\rm orb} = 18.4\,{\rm Myr}$.

At each timestep, we check for all cells whether (1) $\rho>\rho_{\rm LP}$, where we take the threshold $\rho_{\rm LP} \equiv 8.86 c_s^2/(G\Delta x ^2)$ (for $c_s$ the sound speed) as indicative of runaway gravitational collapse, based on evaluation of the Larson-Penston asymptotic profile at radius half of the grid spacing $\Delta x$, (2) $\Phi_{\rm self}$ has a local minimum, and (3) the velocity is converging in all directions.
If the above three conditions are met simultaneously in a given cell, we spawn a sink particle representing a star cluster and convert a portion of the gas mass in the surrounding 27 cells to the mass of the sink particle.
The sink particles are allowed to accrete gas from their neighborhood until the onset of first SN explosion, which occurs at $t\sim 4\,{\rm Myr}$ after creation.
We track the trajectories of the sink particles using the Boris integrator \citep[see also Appendix of \citetalias{moon21}]{boris70} which conserves the Jacobi integral very well.

Assuming that the sink particles fully sample the \citet{kroupa01} initial mass function, we assign the far-ultraviolet (FUV) luminosities based on their mass and age using {\tt STARBURST99} \citep{leitherer99}.
Treating the individual sink particles as  sources of FUV radiation, we set the mean FUV intensity $J_{\rm FUV}$ based on radiation transfer in plane-parallel geometry, with an additional local attenuation in dense regions.
We then scale the photoelectric heating rate $\Gamma_{\rm PE}$ in proportion to $J_{\rm FUV}$, while allowing for the metagalactic FUV background (a small contribution).
The heating rate $\Gamma_{\rm CR}$ by cosmic ray ionization is taken to be proportional to the SFR averaged over a $40\,{\rm Myr}$ timescale, assuming that cosmic rays are accelerated in SN remnants.

Sink particles with age $\sim 4$--$40\,{\rm Myr}$ produce feedback representing type II SNe, with rates based on the tabulation in {\tt STARBURST99}.
The amount of the feedback energy or momentum depends on the gas density. 
If the gas density surrounding an SN is so low that the Sedov-Taylor stage is expected to be resolved, we inject the total energy $E_{\rm SN}=10^{51}\,{\rm erg}$, with $72\%$ and $28\%$ in thermal and kinetic forms, respectively.
If the gas density is instead too high for the shell formation radius to be resolved, we inject the radial momentum of $p_*=2.8\times 10^5\,M_\odot\,{\rm km\,s^{-1}}\,(n_{\rm H}/{\rm cm}^{-3})^{-0.17}$, corresponding to the terminal momentum injected by a single SN \citep{cgkim15}.
Each SN also returns the ejecta mass of $M_{\rm ej} = 10\,M_\odot$ from a sink particle back to the gas.
The reader is referred to \citet{cgkim17} for the full description of the sink formation and SN feedback.

\subsection{Models}

\begin{deluxetable*}{ccccc}
    \tablecaption{Summary of all models\label{tb:models}}
    \tablehead{
\colhead{Model} &
\colhead{$\dot{M}_{\rm in,0}$ ($M_\odot\,{\rm yr}^{-1}$)} &
\colhead{${\cal R}(\dot{M}_{\rm in})$} &
\colhead{$\Delta \tau_{\rm in}$ (${\rm Myr}$)} &
\colhead{Remarks} \\
\colhead{(1)} &
\colhead{(2)} &
\colhead{(3)} &
\colhead{(4)} &
\colhead{(5)}
    }
    \startdata
{\tt constant} & 0.5           & 1  & $\infty$  & constant inflow rate; identical to model {\tt L1} of \citetalias{moon21} \\ 
{\tt P15}      & 0.5           & 20 & 15  & $\dot{M}_{\rm in}$ varying sinusoidally in time \\
{\tt P50}      & 0.5           & 20 & 50  & $\dot{M}_{\rm in}$ varying sinusoidally in time \\
{\tt P100}     & 0.5           & 20 & 100 & $\dot{M}_{\rm in}$ varying sinusoidally in time \\\hline
{\tt asym}     & 0.5           & 1  & -   & 9 times higher inflow rate in the upper nozzle than the lower nozzle \\
{\tt off}      & 1.0 $\to$ 0.5\tablenotemark{a} & 1\tablenotemark{c}  & -   & lower nozzle shut off after $150\,{\rm Myr}$ \\
{\tt boost}    & 0.1 $\to$ 0.5\tablenotemark{b} & 1\tablenotemark{c}  & -   & inflow rate from the upper nozzle boosted after $150\,{\rm Myr}$
    \enddata
    \tablecomments{(1) Model name; (2) Time-averaged mass inflow rate; (3) and (4) Amplitude and period of the  inflow rate variation, respectively;  (5) Comments}
    \tablenotetext{a}{$\dot{M}_{\rm in} = 1\,M_\odot\,{\rm yr}^{-1}$ and $0.5\,M_\odot\,{\rm yr}^{-1}$ before and after $t=150\,{\rm Myr}$, respectively.}
    \tablenotetext{b}{$\dot{M}_{\rm in} = 0.1\,M_\odot\,{\rm yr}^{-1}$ and $0.5\,M_\odot\,{\rm yr}^{-1}$ before and after $t=150\,{\rm Myr}$, respectively.}
    \tablenotetext{c}{The inflow rate of models {\tt off} and {\tt boost} is constant except for a discontinuous jump at $t=150\,{\rm Myr}$.}
\end{deluxetable*}

In our semi-global models, the simulation domain is initially filled with rarefied gas with number density $n_{\rm H} = 10^{-5}\,\exp\left[{-|z|/(50\,{\rm pc})}\right]\,{\rm cm^{-3}}$ and temperature $T=2\times 10^4\,\rm K$.
Gas flows in to the domain through two nozzles located at the $y$-boundaries, mimicking bar-driven gas inflows along dust lanes.
The mass inflow rate $\dot{M}_{\rm in}$ is controlled by varying the gas density inside the nozzles.
\citetalias{moon21} focused on the cases with constant $\dot{M}_{\rm in}$ over time.
In this paper we report on two additional suites of models to explore the effects of time variations and asymmetry in the mass inflow rates.
Table \ref{tb:models} lists the models we consider.
 
In the first suite, we let the mass inflow rate vary sinusoidally with time in logarithmic scale as 
\begin{equation}\label{eq:inflow_rate}
    \ln\left(\frac{\dot{M}_{\rm in}}{M_\odot\,{\rm yr}^{-1}}\right) = A - B\cos\left(\frac{2\pi t}{\Delta \tau_{\rm in}}\right),
\end{equation}
where $\Delta \tau_{\rm in}$ is the oscillation period.
The dimensionless constants $A$ and $B$ parameterize the time-averaged inflow rate $\dot{M}_{\rm in,0}\equiv \int_0^{\Delta \tau_\text{in}} \dot{M}_\text{in} dt /\int_0^{\Delta \tau_\text{in}} dt$ and the oscillation amplitude ${\cal R}(\dot{M}_{\rm in}) \equiv \max(\dot{M}_{\rm in})/\min (\dot{M}_{\rm in})$, such that $\dot{M}_{\rm in,0}/(M_\odot\,{\rm yr}^{-1})=I_0(B)e^A$ and ${\cal R}(\dot{M}_{\rm in})=e^{2B}$, where $I_0$ is the zeroth-order modified Bessel function of the first kind.
In the first suite, we fix $\dot{M}_{\rm in,0} = 0.5\,M_\odot\,{\rm yr}^{-1}$ and ${\cal R}(\dot{M}_{\rm in}) = 20$, corresponding to $A=-1.19$ and $B=1.50$, and consider three models {\tt P15}, {\tt P50}, and {\tt P100} with $\Delta \tau_{\rm in}=15$, $50$, and $100\,{\rm Myr}$, respectively.
We also include model {\tt constant} with ${\cal R}(\dot{M}_{\rm in}) = 1$ and $\Delta \tau_\text{in}=\infty$, which is identical to model {\tt L1} in \citetalias{moon21}.
Note that $\dot{M}_{\rm in}(t)$ denotes the total inflow rate through the two nozzles: the mass inflow rate from one nozzle is $\dot{M}_{\rm in}(t)/2$ in the first suite. 

In the second suite, we consider three models termed {\tt asym}, {\tt off}, and {\tt boost} to investigate the effect of asymmetric inflows.
In model {\tt asym}, the mass inflow rates from the upper (at the positive $y$-boundary) and the lower (at the negative $y$-boundary) nozzles are $0.45\,M_\odot\,{\rm yr}^{-1}$ and $0.05\,M_\odot\,{\rm yr}^{-1}$, respectively, and do not vary in time throughout the simulation.
The mass inflow rate in model {\tt off} is initially $0.5\,M_\odot\,{\rm yr}^{-1}$ from each nozzle, but the lower nozzle is abruptly turned off at $t = 150\,{\rm Myr}$, while the upper nozzle keeps supplying the gas with the same rate all the time. 
In model {\tt boost}, the mass inflow rate from each nozzle is set to $0.05\,M_\odot\,{\rm yr}^{-1}$ at early time, but the upper nozzle suddenly boosts the inflow rate to $0.45\,M_\odot\,{\rm yr}^{-1}$ at $t = 150\,{\rm Myr}$, while the inflow rate from the lower nozzle is kept unchanged. 
Note that the total mass inflow rate for all three models in the second suite is $0.5\,M_\odot\,{\rm yr}^{-1}$ after $t=150\,{\rm Myr}$.

\section{Results}\label{s:results}

In this section, we present the temporal histories of the SFR, gas mass, and depletion time, and examine the relation between the SFR and the inflow rate. We also present the results of the second suite of models with asymmetric inflows in terms of the spatial distributions of young clusters. 

\subsection{Time Variation of the SFR}

\begin{figure}
    \includegraphics[width=\linewidth]{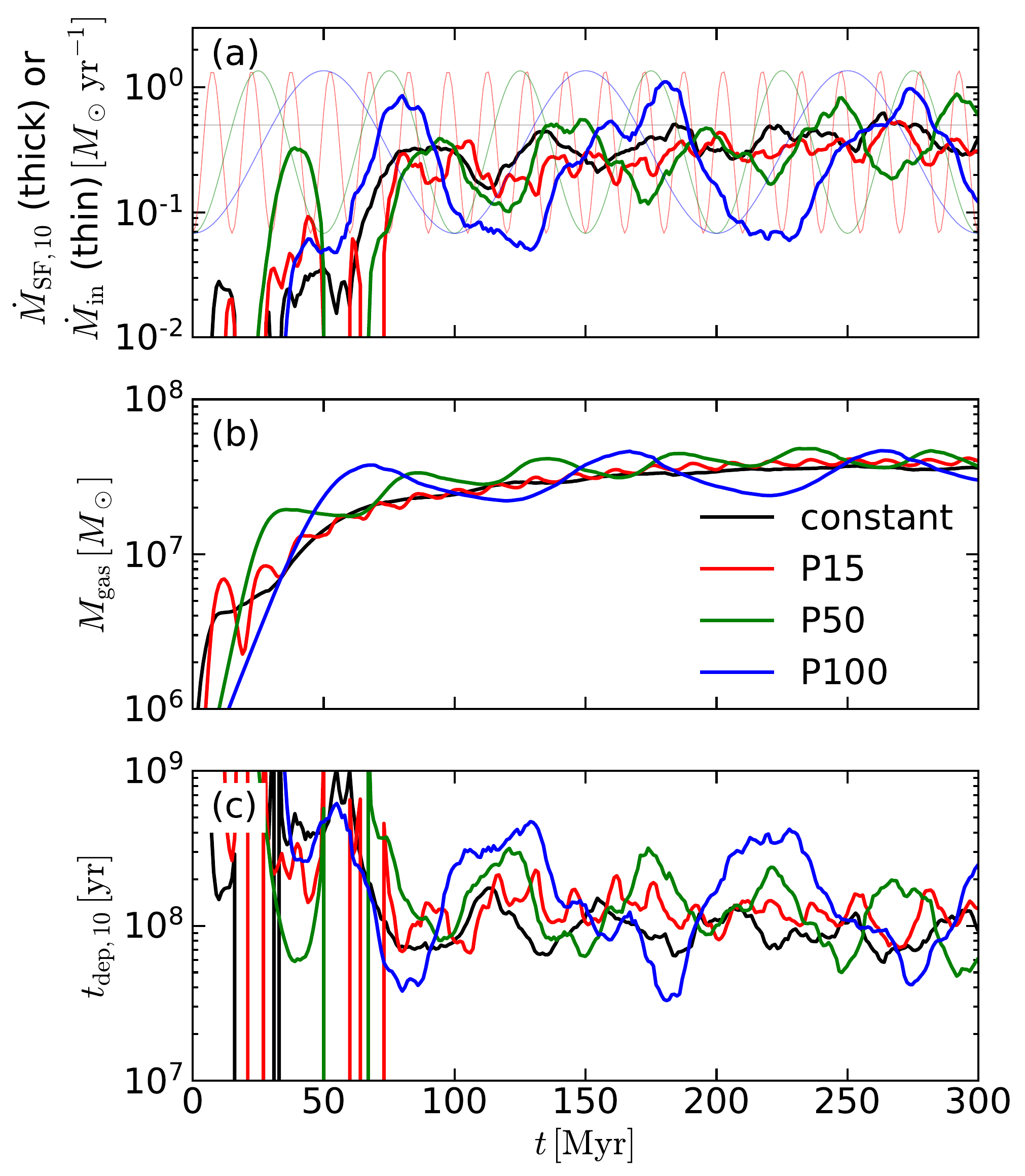}
    \caption{Temporal histories of ($a$) the SFR (thick) and the inflow rate (thin), ($b$) the total gas mass inside the computational domain, and ($c$) the gas depletion time for models {\tt constant} (black), {\tt P15} (red), {\tt P50} (green), and {\tt P100} (blue).}
    \label{fig:history}
\end{figure}

We calculate the SFR using the sink particles that formed in the past $10\,{\rm Myr}$ as 
\begin{equation}\label{eq:SFR}
    \dot{M}_\text{SF,10}(t) = \frac{M_{\rm sp}(t)-M_{\rm sp}(t-10\,{\rm Myr})}{10\,{\rm Myr}},
\end{equation}
where $M_{\rm sp}(t)$ is the total mass in the sink particles at time $t$.
Figure \ref{fig:history} plots for models {\tt constant}, {\tt P15}, {\tt P50}, and {\tt P100} the temporal histories of the SFR, the total gas mass $M_{\rm gas}$ in the computational domain, and the depletion time averaged over the past $10\,{\rm Myr}$,
\begin{equation}\label{eq:tdep10}
    t_\text{dep,10}\equiv M_{\rm gas}/\dot{M}_\text{SF,10}.
\end{equation}
We note that a depletion time  $t_\text{dep,40}$ over $40\,{\rm Myr}$ (or other time) can be analogously defined.
At early time, the gas streams from the nozzles collide with each other multiple times, increasing the SFR temporarily.
A nuclear ring forms at $t\sim 100\,{\rm Myr}$, after which the evolution depends strongly on $\Delta \tau_\text{in}$.
While models {\tt constant} and {\tt P15} show weak fluctuations, the SFR in models {\tt P50} and {\tt P100} varies quasi-periodically with large amplitudes, even though the gas mass is almost the same.
In all models with $\dot{M}_\text{in,0}=0.5\,M_\odot\,{\rm yr}^{-1}$, the ring mass is $M_\text{gas}\sim 4 \times 10^7 M_\odot$, weakly varying with time, with peaks in the mass phase-delayed relative to the peak in $\dot{M}_\text{in}$.
For quantitative comparison, we define the fluctuation amplitude ${\cal R}$ by taking the ratio of the maximum to minimum values of $\dot{M}_\text{SF,10}$, $M_{\rm gas}$, and $t_\text{dep,10}$ during $t=200$--$300\,{\rm Myr}$.
Table \ref{tb:results} gives ${\cal R}(\dot{M}_\text{SF,10})$, ${\cal R}(M_{\rm gas})$, and ${\cal R}(t_\text{dep,10})$ for all the models. 

Although the mass inflow rate for model {\tt constant} is constant in time, its SFR shows random fluctuations with amplitude ${\cal R}(\dot{M}_\text{SF,10})=2.4$, due to the turbulence driven by the SN feedback, consistent with the results of \citetalias{moon21}.
When the mass inflow rate oscillates with time, the fluctuation amplitudes increase with increasing $\Delta \tau_{\rm in}$.
In model {\tt P100}, for instance, the SFR varies by a factor of 17 while the gas mass varies by a factor of 2, resulting in a factor of 10 variations of the depletion time.
The fluctuations become smaller in model {\tt P50}, with a factor of 5 variations in the SFR and the depletion time.
When the inflow rate oscillates very rapidly as in model {\tt P15}, the fluctuation amplitudes are almost the same as those in model {\tt constant}, as well as in the second suites of models {\tt asym}, {\tt off}, and {\tt boost} in which the mass inflow rate is constant after $t=150\,{\rm Myr}$.

\begin{deluxetable}{lccc}
    \tablecaption{Fluctuation amplitudes of $\dot{M}_\text{SF,10}$, $M_{\rm gas}$, and $t_\text{dep,10}$ \label{tb:results}}
    \tablehead{
\colhead{Model} &
\colhead{${\cal R}(\dot{M}_\text{SF,10})$} &
\colhead{${\cal R}(M_{\rm gas})$} &
\colhead{${\cal R}(t_\text{dep,10})$}
    }
    \startdata
{\tt constant} & 2.4 & 1.1 & 2.3  \\
{\tt P15}      & 2.3 & 1.2 & 2.4  \\
{\tt P50}      & 5.4 & 1.3 & 5.2  \\
{\tt P100}     & 17  & 2.0 & 10   \\\hline
{\tt asym}     & 2.3 & 1.1 & 2.4  \\
{\tt off}      & 2.5 & 1.1 & 2.8  \\
{\tt boost}    & 2.4 & 1.4 & 2.6
    \enddata
\end{deluxetable}

\subsection{Relation between the SFR and the Inflow Rate}

\begin{figure}
    \includegraphics[width=\linewidth]{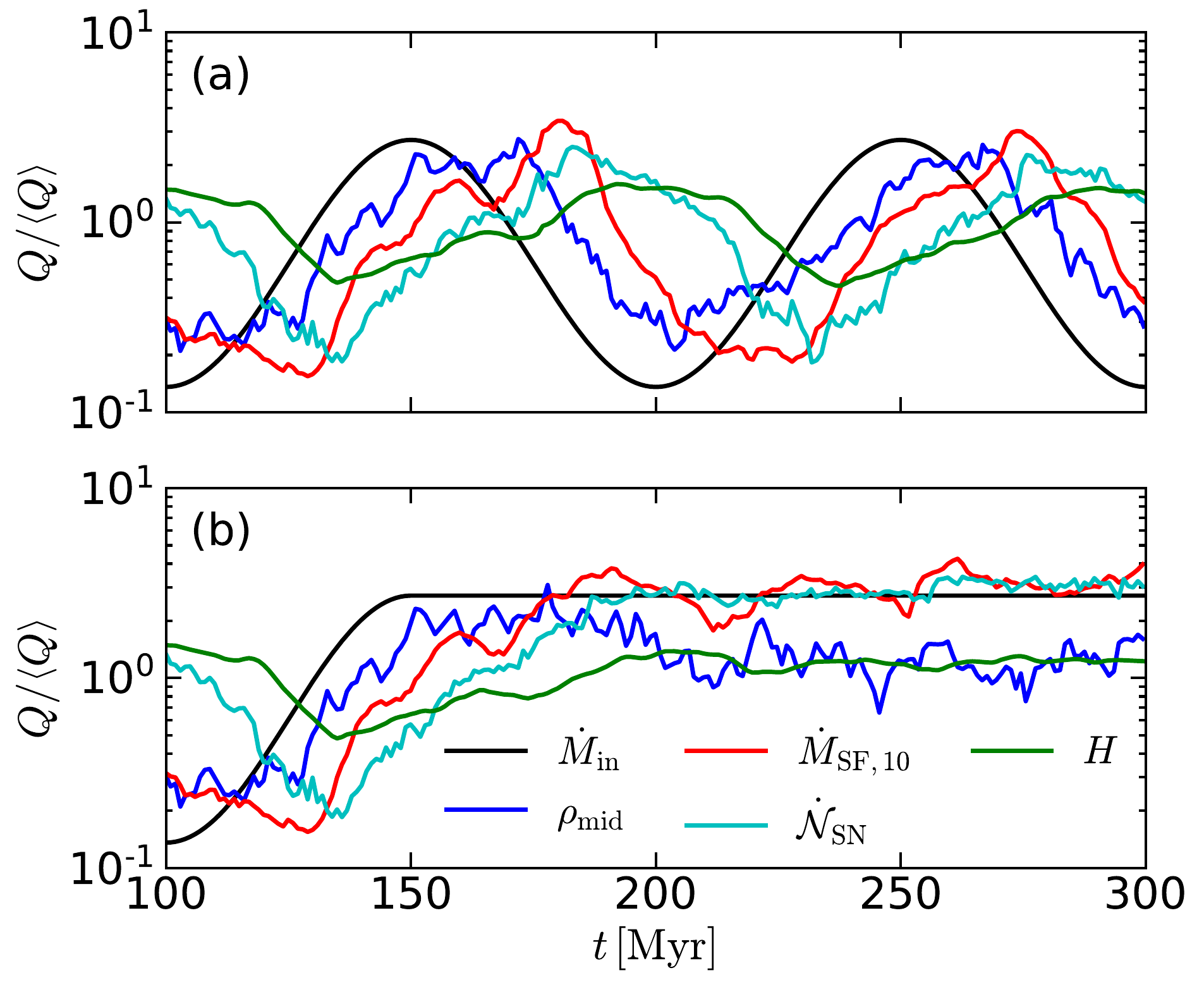}
    \caption{Time histories of the inflow rate (black), the mean midplane density of cold-warm gas with temperature $<2\times 10^4\,{\rm K}$ (blue), the SFR (red), the rate of SN explosions (cyan), and the gas scale height (green) for model {\tt P100}. Because the SN feedback is active for $4$--$40\,{\rm Myr}$ after the star formation, the SN rate lags behind the SFR. Panels ($a$) and $(b)$ correspond, respectively, to model {\tt P100} and its restart experiment from $t=150\,{\rm Myr}$, with $\dot{M}_\text{in}$ fixed to the maximum rate (see text). All quantities are normalized by the time-averaged values between $t=100$--$300\,{\rm Myr}$ of model {\tt P100} for comparison.}
    \label{fig:exam_history}
\end{figure}

Figure \ref{fig:history}($a$) shows that for models {\tt P100} and {\tt P50}, the SFR varies coherently with the inflow rate with some time delay.
To quantify the delay, we define the characteristic delay time $t_{\rm delay}$ as the time lag at which the cross-correlation between the SFR and the inflow rate is maximized, that is, $t_{\rm delay}\equiv \argmax_\tau (\dot{M}_\text{SF,10}\star \dot{M}_{\rm in})(\tau)$.
Our numerical results correspond to  $t_{\rm delay} = 25\,{\rm Myr}$ and $t_{\rm delay} = 19\,{\rm Myr}$ for models {\tt P100} and {\tt P50}, respectively.
Multiple effects contribute to the time delay.
First, it takes approximately $3\,{\rm Myr}$ for the inflowing gas to travel from the nozzle to the ring located at $R_\text{ring}$.
Second, our SFR measurement introduces some delay, approximately $4\,{\rm Myr}$, because it counts all star formation events in the past $10\,{\rm Myr}$.
Subtracting the above two effects, the delay time for models {\tt P100} and {\tt P50} reduces to $18\,{\rm Myr}$ and $12\,{\rm Myr}$, respectively.
The resulting delay time can be interpreted as the timescale for the newly accreted gas first to enhance the overall mass in the ring, and then to produce local enhancements in the density above the (numerical) threshold for star formation.  We note, from Figure \ref{fig:history}($b$), that there are localized peaks in the ring gas mass that are earlier in phase by $\sim 10\,{\rm Myr}$ relative to the peak in SFR for both models {\tt P100} and {\tt P50}. This phase offset of the peak in $M_\text{gas}$ is comparable to $\sim 0.7$ vertical oscillation periods associated with $\Phi_\text{ext}$. This appears to be the minimum time required for an enhancement in the SFR to develop via stochastic processes such as turbulent compression and gravitational contraction.
 
Unlike in models {\tt P100} and {\tt P50}, there is no apparent correlation between the SFR and the inflow rate in model {\tt P15}.
This is because the SFR fluctuations due to the time variations of $\dot{M}_\text{in}$ are too weak to stand out against the feedback-driven fluctuations (see below).

To better understand what drives temporal variations of the SFR, Figure \ref{fig:exam_history}($a$) plots for model {\tt P100} the temporal histories of the mass inflow rate, the mean midplane density $\rho_\text{mid}$ of cold-warm gas with $T<2\times 10^4\,{\rm K}$, the SFR, the number of SN explosions per unit time $\dot{\mathcal{N}}_{\rm SN}$, and the gas scale height $H\equiv \left(\int \rho z^2\,dV/\int \rho\,dV \right)^{1/2}$.
All the quantities are normalized by their respective time average over $t=100$--$300\,{\rm Myr}$.
Overall, the time variations of $\dot{M}_\text{in}$ lead to the changes in, sequentially, $\rho_\text{mid}$, SFR, $\dot{\mathcal{N}}_{\rm SN}$, and $H$, with approximate delay times of $12$, $25$, $36$, and $47\,{\rm Myr}$, respectively.
This makes sense since the mass inflows first enhance the ring density to promote star formation.
The associated enhancement in SN feedback then inflates the ring vertically, increasing $H$.
Both the mass inflows and feedback can affect the SFR by changing $\rho_\text{mid}$, with the latter being through $H$.
All the quantities vary quasi-periodically with the dominant period of $100\,{\rm Myr}$.

To examine whether the quasi-periodic cycles of the SFR are really driven by $\dot{M}_\text{in}(t)$ rather than the SN feedback, we restart model {\tt P100} from $t=150\,{\rm Myr}$ by fixing the inflow rate to $\text{max}(\dot{M}_\text{in})=1.36\,M_\odot\,\text{yr}^{-1}$ thereafter, which we term model {\tt P100\char`_restart}.
Figure \ref{fig:exam_history}($b$) plots the resulting time histories of $\dot{M}_\text{in}$, $\rho_\text{mid}$, SFR, $\dot{\mathcal{N}}_\text{SN}$, and $H$ of model {\tt P100\char`_restart}, normalized by the time-averaged values of model {\tt P100} for direct comparison.
With fixed $\dot{M}_\text{in}$, the system reaches a quasi-steady state at $t\sim 200\,{\rm Myr}$ in which the SFR and the other quantities do not vary much with time.
The short-term ($\lesssim 40\,{\rm Myr}$) fluctuations in the averaged quantities are due purely to turbulence driven by the SN feedback, which is active for $4$--$40\,{\rm Myr}$ after the star formation.
The corresponding fluctuation amplitude in the SFR is ${\cal R}(\dot{M}_\text{SF,10})=2.4$ in model {\tt P100\char`_restart}, which is 7 times smaller than that in model {\tt P100}, but comparable to  the fluctuation amplitude in model {\tt constant}.
This demonstrates that the large-amplitude, quasi-periodic variations of the SFR in model {\tt P100} are caused by $\dot{M}_\text{in} (t)$, while the stochastic SN feedback is responsible for small-amplitude (a factor of $\sim 2$), short-term fluctuations of the SFR with timescale $\lesssim 40\,{\rm Myr}$.
Of course, longer-term variations in the SN feedback rate {\it induced} by variations in the inflow rate and SFR are also dynamically important, as noted above (see also Section \ref{sec:self-reg})

\begin{figure}
    \includegraphics[width=\linewidth]{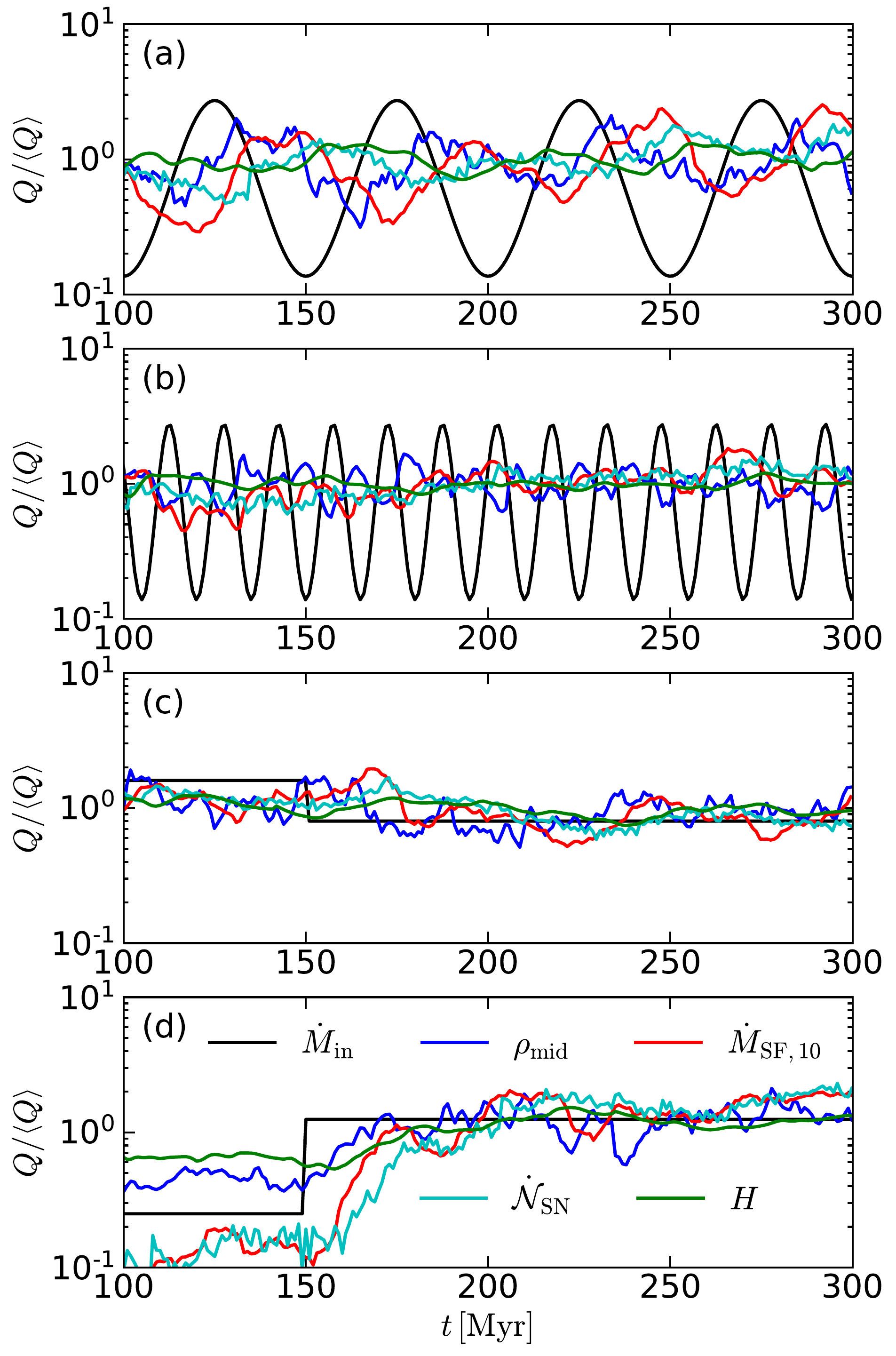}
    \caption{Similar to Figure \ref{fig:exam_history}, for models  ($a$) {\tt P50}, ($b$) {\tt P15}, ($c$) {\tt off}, and ($d$) {\tt boost}.} 
    \label{fig:exam_history_others}
\end{figure}

Figure \ref{fig:exam_history_others} plots the histories of the various quantities for models {\tt P50}, {\tt P15}, {\tt off}, and {\tt boost}. Model {\tt P50} behaves qualitatively similarly to model {\tt P100}, in that the oscillating inflows drive long-term ($\sim50\,\rm Myr$), large-amplitude variations of the SFR, while short-term, small-amplitude fluctuations are due to the SN feedback.
Note however that the oscillation amplitudes of the SFR decreases with decreasing $\Delta \tau_\text{in}$. This is because in our simulations about 80\% of the inflowing gas turns to stars \citepalias{moon21}, so that neglecting the effect of the SN feedback changing the scale height, the SFR is roughly given by 
\begin{equation}
    \dot{M}_\text{SF,10}(t) \approx  \frac{0.8}{10\rm\,Myr}\int_{t-t_\text{delay}-10\rm\,Myr}^{t-t_\text{delay}} \dot{M}_\text{in}(t)dt,
\end{equation}
where $t_\text{delay}$ is the delay time between the mass inflow and star formation mentioned above.
Since our SFR is proportional to the gas mass accreted over a $10\rm\,Myr$ interval, its amplitude should be an increasing function of $\Delta \tau_\text{in}$ even though the oscillation amplitude of $\dot{M}_\text{in}$ is the same.
Note that the averaging interval of $10\,\rm Myr$ is likely a lower limit, considering that the time taken for an inflowing gas parcel to turn into stars is not unique but distributed around $t_{\rm delay}$.
When $\Delta \tau_\text{in} \lesssim 10\rm\,Myr$, the effect of the temporal variations of $\dot{M}_\text{in}$ on the SFR would be smoothed out almost completely.
In addition, small-amplitude fluctuations of the SN feedback are present in all models, tending to reduce the effect of the time-varying inflow rate for small $\Delta \tau_\text{in}$.
Considering the SN timescale of $\sim 40\,{\rm Myr}$ 
(combined with the short vertical dynamical  time in galactic center regions)
it is very likely that the star formation can follow the variations imposed by the inflow when $\Delta \tau_\text{in}\gtrsim 40\,{\rm Myr}$, as is indeed seen in models {\tt P50} and {\tt P100}.
In model {\tt off} or {\tt boost}, where the inflow rate drops by a factor of 2 or increases by a factor of 5, respectively, at $t=150\,{\rm Myr}$, the SFR decreases or increases by a similar factor and undergoes feedback-induced, small-amplitude fluctuations with ${\cal R}(\dot{M}_\text{SF,10}) = 2.5$, similarly to that in model {\tt constant}.

\subsection{Self-regulation Theory}\label{sec:self-reg}

\begin{figure}
    \plotone{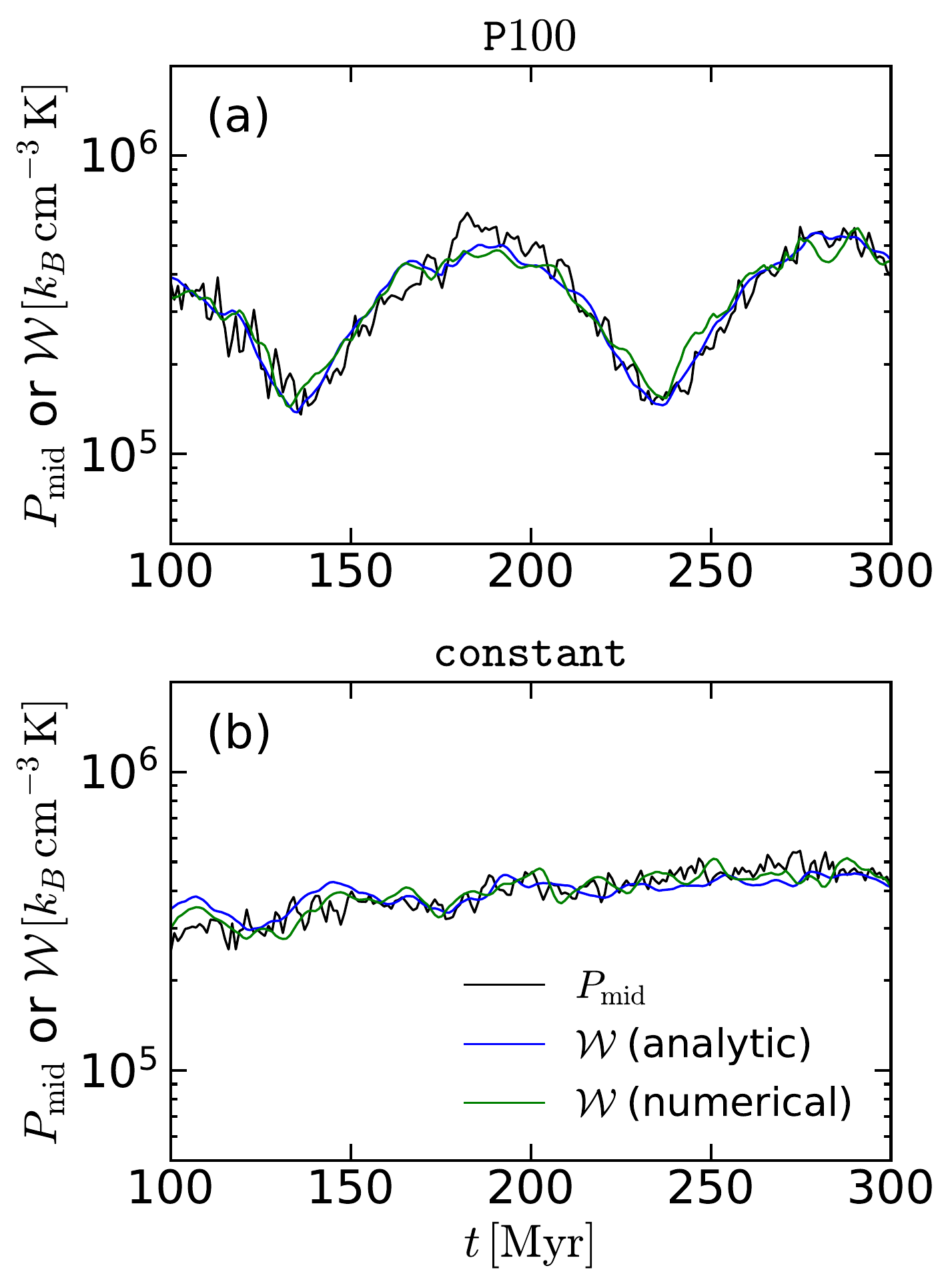}
    \caption{Temporal evolution of midplane pressure $P_\text{mid}$ and gas weight $\mathcal{W}$ for models $(a)$ {\tt P100} and $(b)$ {\tt constant}. The blue and green lines correspond to the weights calculated by using the analytic and numerical $\left<g_z\right>$, respectively (see text). In all epochs, 
    $\mathcal{W}\approx P_\text{mid}$, indicating that the system is approximately in instantaneous vertical equilibrium.}
    \label{fig:equilibrium}
\end{figure}

As Figure \ref{fig:history} shows, the gas depletion time in model {\tt P100} varies by about an order of magnitude, while the ring gas mass is almost constant. In this subsection, we shall show that this is consistent with the results of the self-regulated star formation theory \citep{ostriker10,ostriker11} provided that the time delay between star formation and the ensuing SN feedback is properly considered (see Figure \ref{fig:exam_history}$a$).

According to the self-regulation theory, an (unmagnetized) disk in vertical dynamical equilibrium should obey
\begin{equation}\label{eq:equilibrium}
    \begin{split}
        P_{\rm mid} &\equiv \frac{1}{A_{\rm ring}}\iint (P + \rho v_z^2)|_{z=0}\,dxdy\\
        &= \mathcal{W} \equiv \frac{1}{2}\Sigma_\text{gas}\left<g_z\right>, 
  \end{split}
\end{equation}
where $P_{\rm mid}$ is the total (thermal plus turbulent) midplane pressure averaged over the ring area $A_{\rm ring}$ (see below), $\mathcal{W}$ is the weight of the overlying gas above (or below) the midplane, $\Sigma_\text{gas}$ is the mean gas surface density within the ring area, and $\left<g_z\right>$ is the density-weighted mean vertical gravity defined as
\begin{equation}\label{eq:gz_def}
    \left<g_z\right> \equiv \frac{\iiint \rho g_z\,dxdydz}{\iiint \rho \,dxdydz}.
\end{equation}
Here, the integration is performed over the annular region between $R_{\rm min}=400\,\text{pc}$ and $R_{\rm max} = 800\,{\rm pc}$, where $A_{\rm ring} = \pi(R_{\rm max}^2-R_{\rm min}^2)$, and over the upper (or lower) half of the computational domain in the vertical direction.
Equation \eqref{eq:equilibrium} assumes that the pressure above the gas layer is small compared to $P_\mathrm{mid}$, i.e. $\Delta (P+\rho v_z^2) \approx P_\mathrm{mid}$.
When the gravitational potential is dominated by the stellar bulge (Equation \ref{eq:bulge}), it can be shown that
\begin{equation}\label{eq:gz_proxy}
    \left<g_z\right> = f\left(\frac{2}{\pi}\right)^{1/2} \frac{GM_bH}{R_{\rm ring}^3},
\end{equation}
where $f$ is a dimensionless parameter of order unity that depends on the spatial gas distribution: e.g., 
$f=1$ for a thin ring with a Gaussian density distribution in the vertical direction. We find $f=0.64$ and $0.72$ for model {\tt P100} and {\tt constant}.
Figure \ref{fig:equilibrium} plots the relationship between $P_\text{mid}$ and $\mathcal{W}$ for models {\tt P100} and {\tt constant}, showing that the two quantities agree with each other within $\sim12\%$ and $\sim 8\%$, respectively: other models also show a good agreement between $P_\text{mid}$ and $\mathcal{W}$.
This  demonstrates that the system maintains an instantaneous vertical equilibrium, while undergoing (quasi-periodic) long-term oscillations in response to changes in the inflow rate.
We also note that the weights from the analytic (Equation \ref{eq:gz_proxy}) and numerically measured $\left<g_z\right>$ (Equation \ref{eq:gz_def}) are practically the same, indicating that Equation \eqref{eq:gz_proxy} is a good approximation for the vertical gravitational field in our simulations.

\begin{figure}
    \includegraphics[width=\linewidth]{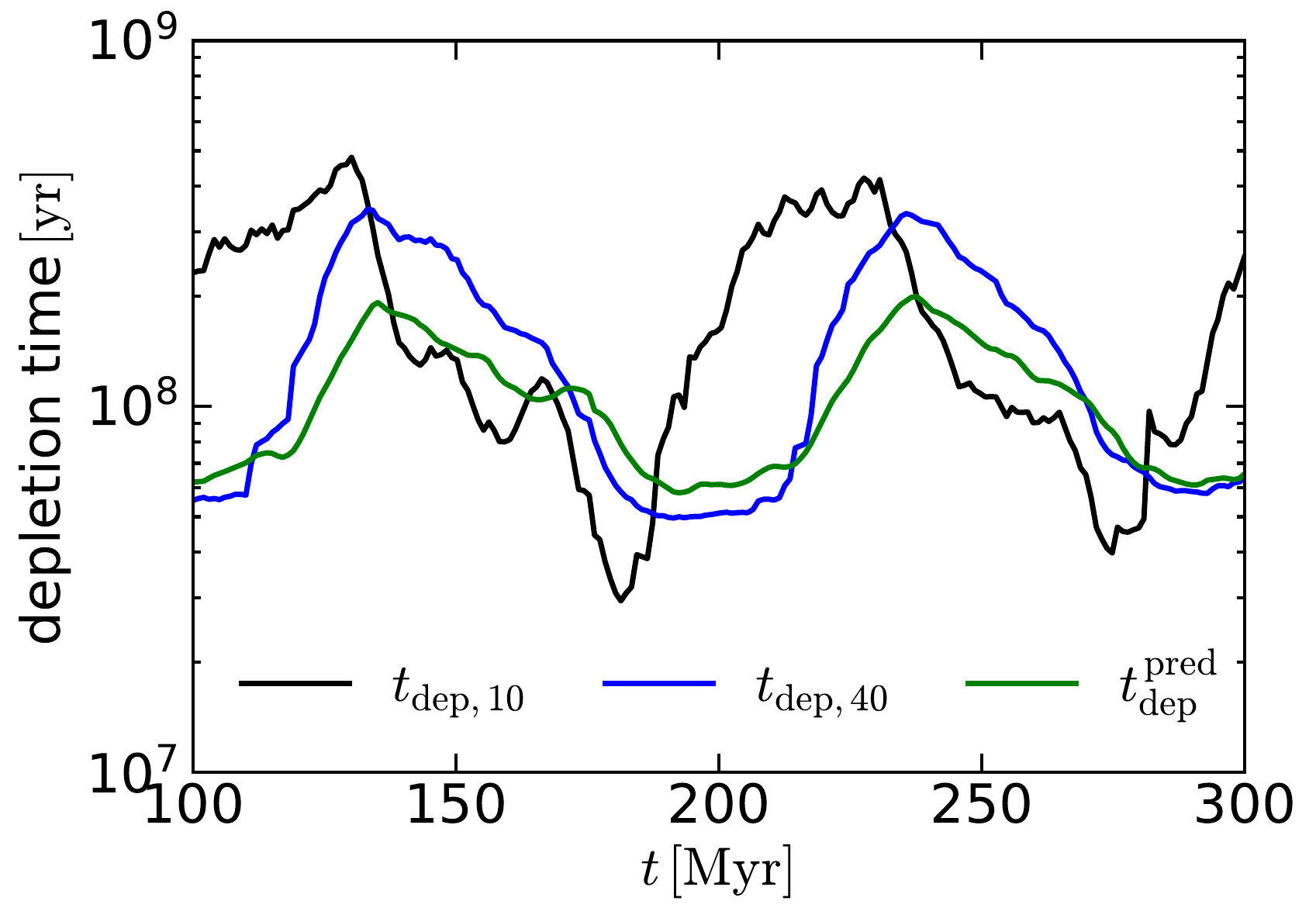}
    \caption{Comparison for model {\tt P100} of the measured depletion times averaged over $10\,\text{Myr}$ ($t_\text{dep,10}$, black) and $40\,\text{Myr}$ ($t_\text{dep,40}$, blue), together with the predicted depletion time (Equation \ref{eq:tdep_pred}, green). Note that $t_\text{dep}^\text{pred}$ follows $t_\text{dep,40}$ quite well.}
    \label{fig:tdep}
\end{figure}

The self-regulation theory further asserts that the midplane pressure is sustained by feedback. In the present case, SN feedback supplies both thermal (through hot bubbles) and turbulent pressures, although other forms of feedback or other pressure contributions (e.g., magnetic pressure) may be important more generally \citep{ostriker10,ostriker11,cgkim15b}.
Assuming that the midplane pressure is proportional to the SFR surface density $\Sigma_\text{SFR}\equiv \dot{M}_{\rm SF}/A_{\rm ring}$, one can write $P_{\rm mid} = \Upsilon_{\rm tot}\Sigma_\text{SFR}$, where $\Upsilon_{\rm tot}$ is the total feedback yield \citepalias[see also, \citealt{cgkim11,cgkim13,cgkim15b}]{moon21}.
We find $\Upsilon_{\rm tot} = 340\,{\rm km\,s^{-1}}$ from the time averaged $P_{\rm mid}$ and $\Sigma_\text{SFR}$ for model {\tt P100} (with $\sim 2\%$ differences for other models).
Identifying the pressure predicted from dynamical equilibrium  (Equation \ref{eq:equilibrium}) with the pressure predicted from star formation feedback then gives the prediction for the gas depletion time (Equation \ref{eq:tdep10}) as 
\begin{equation}\label{eq:tdep_pred}
    t_\text{dep}^\text{pred} = \frac{2\Upsilon_{\rm tot}}{\left<g_z\right>}
    =\frac{(2\pi)^{1/2}R_{\rm ring}^3\Upsilon_{\rm tot}}{fGM_bH}.
\end{equation}
Figure \ref{fig:tdep} compares the measured $t_\text{dep,10}$ and the predicted $t_\text{dep}^\text{pred}$ for model {\tt P100}.
Although $t_\text{dep}^\text{pred}$ oscillates in time in a similar fashion to the directly-measured $t_\text{dep,10}$ as the ring repeatedly shrinks and expands vertically to change $H$, the amplitude and phase are not well matched.

The temporal offset between $t_\text{dep,10}$ and $t_\text{dep}^\text{pred}$ implies that one has to consider the time delay between star formation and SN feedback: while $\dot{M}_\text{SF,10}$ and $t_\text{dep,10}$ only accounts for the stars formed in the past $10\,\text{Myr}$, the SN feedback that sustains the midplane pressure also depends on previous star formation, occuring in star particles with age up to $40\,\text{Myr}$ \citep{leitherer99,cgkim17}.
Since the scale height $H$ (and $P_\text{mid}$, implicitly) in Equation \eqref{eq:tdep_pred} is responsive to SNe (Figure \ref{fig:exam_history}$a$), the corresponding predicted depletion time is sensitive to a longer-term average of the SFR.
This motivates us to compare $t_\text{dep}^\text{pred}$ with the depletion time $t_\text{dep,40}$ averaged over $40\,\text{Myr}$ instead of $t_\text{dep,10}$.
Figure \ref{fig:tdep} shows $t_\text{dep,40}$ agrees with $t_\text{dep}^\text{pred}$ much better than $t_\text{dep,10}$, indicating that the self-regulation theory predicts the {\it time-varying} depletion time averaged over the timescale associated with the dominant feedback process, which is $\sim 40\,\text{Myr}$ for SNe in our current models.
The slight mismatch in the amplitude is due to the secondary effect of time-varying $\Upsilon_\text{tot}$, which we do not consider in this work.
The phase offset and larger fluctuation of $t_\text{dep,10}$ compared to $t_\text{dep,40}$ is because the SN feedback is delayed behind $\dot{M}_\text{SF,10}$.
We note that Equation \eqref{eq:equilibrium} implies $M_\text{gas} = \Sigma_\text{gas} A_\text{ring} \propto P_\text{mid}/\left<g_z\right> \propto \dot{\mathcal{N}}_\text{SN}/H$,
which is roughly constant since $H$ is correlated with $\dot{\mathcal{N}}_{\text SN}$ (see Figure \ref{fig:exam_history}$a$), in agreement with Figure \ref{fig:history}($b$)\footnote{Since most of the gas in our simulations is contained in the ring region, $M_\text{gas}$ agrees with $\Sigma_\text{gas}A_\text{ring}$ within $\sim 8\%$.}.
The above analyses suggest that the self-regulation theory is applicable even when the ring star formation is time-varying, as long as feedback time delays and appropriate temporal averaging windows are taken into account.

\subsection{Spatial Distributions of Star Clusters}
\begin{figure*}
    \plotone{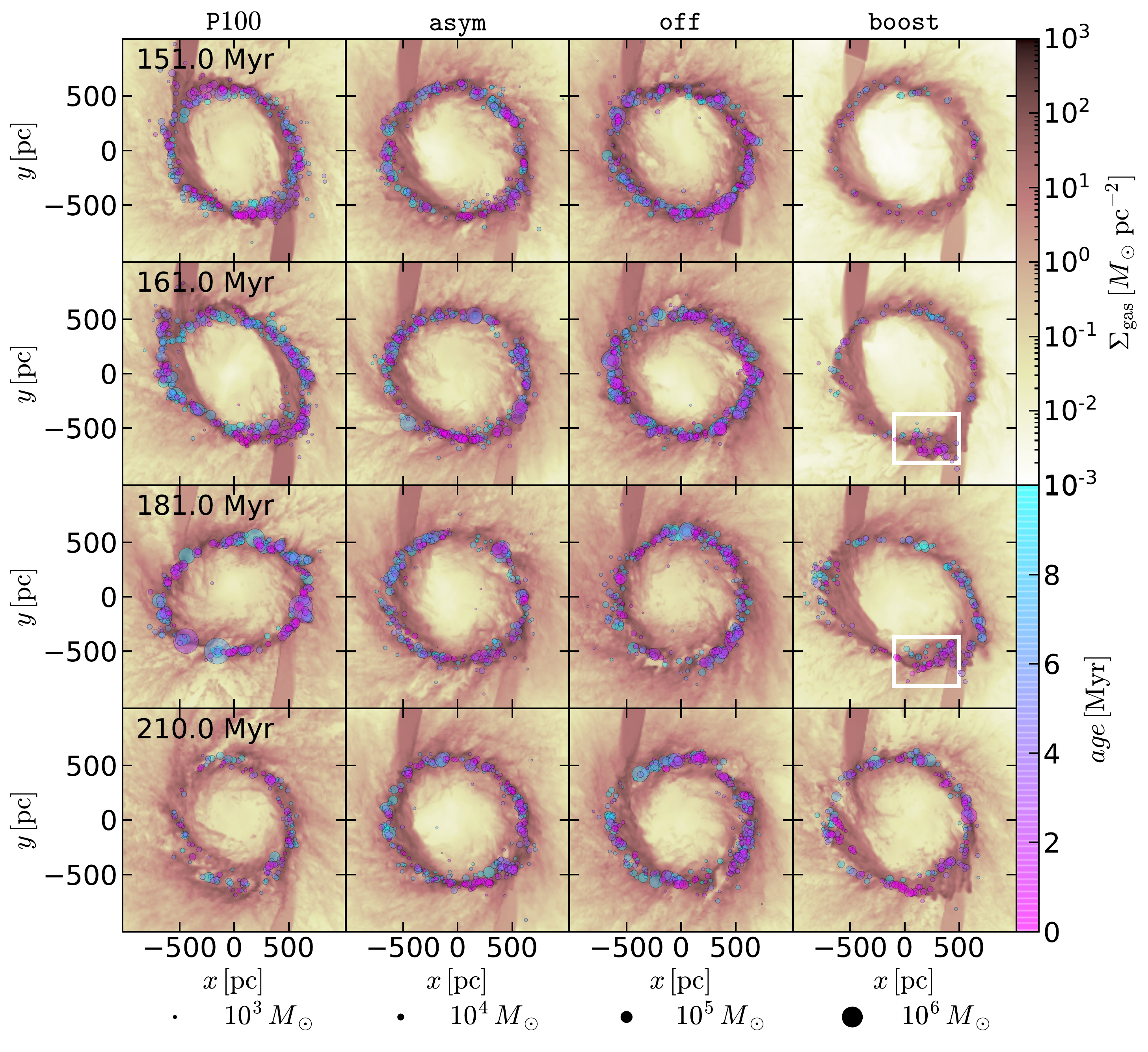}
    \caption{Spatial distributions of the gas surface density and star clusters younger than $10\,{\rm Myr}$, with the color and size coded by their age and mass, respectively, projected on the $x$--$y$ plane for various models. From left to right, each column corresponds to models {\tt P100}, {\tt asym}, {\tt off}, and {\tt boost}. From top to bottom, each row corresponds to the snapshots at $t=151$, $163$, $183$, and $203\,{\rm Myr}$. The white boxes in the two middle panels of model {\tt boost} indicate the star forming regions triggered by the enhanced inflows from the top nozzle at $t=150\,{\rm Myr}$.}
    \label{fig:asym}
\end{figure*}
We now explore how asymmetry in the mass inflows affects the spatial distributions of star particles in the rings.
Figure \ref{fig:asym} plots the projected distributions in the $x$--$y$ plane of gas and young star clusters with age younger than $10\,{\rm Myr}$, for models {\tt P100}, {\tt asym}, {\tt off}, and {\tt boost} from left to right at four selected epochs from $t=151\,{\rm Myr}$ to $210\,{\rm Myr}$. 
Note that the inflow rate in models {\tt off} and {\tt boost} changes abruptly  at $t=150\,{\rm Myr}$.
In model {\tt P100}, the fading gas streams over time manifest the continuous decrease of the inflow rate during this time span.
The time-varying kinetic energy of the streams makes the ring more eccentric and promotes its precession.
With the symmetric mass inflow rate in this model, young star clusters are distributed more-or-less uniformly across the whole length of the ring.

The ring of model {\tt asym} is more circular since the inflowing gas has constant kinetic energy.
While the inflow rate from the upper nozzle is $9$ times higher than that from the lower one in this model, star clusters with age $\lesssim 10\,{\rm Myr}$ are still distributed almost uniformly throughout the ring, as in model {\tt P100}.
This is presumably because the depletion time ($\sim 100\,{\rm Myr}$) is longer than the ring orbital time ($\sim 20\,{\rm Myr}$), allowing the gas from the upper and lower streams to be well mixed before turning into stars.
The inflow rate in model {\tt off} becomes asymmetric after $t=150\,{\rm Myr}$, due to the cessation of the lower stream. Nevertheless, the continued inflow from the upper nozzle smoothly lands on the ring without causing large deformation of the ring. 
The distribution of star particles in model {\tt off} is relatively symmetric despite the asymmetric inflow rate, similarly to model {\tt asym}.

Unlike the other models, however, model {\tt boost} shows lopsided distributions of star particles for a few tens of ${\rm Myr}$ after the boosted inflow, due mainly to the enhanced SFR in the lower part of the ring marked by the white boxes in the second and third rows of the last column in Figure \ref{fig:asym}.
In this model, the boosted inflow from the upper nozzle has such large inertia that it is almost unhindered when it hits the ring at $(x, y) \sim (-500, 400)\,{\rm pc}$ on a nearly ballistic orbit.
The inflowing streams converge and collide with the ring at the opposite side, triggering star formation at $(x, y) \sim (0, -600)\,{\rm pc}$.
This star formation, induced directly by the boosted inflow, makes the overall distribution of star particles lopsided.
However, this phase of asymmetric star formation persists only for a few orbital periods as the ring gradually adjusts its shape and size corresponding to the boosted inflow.
Gas from the boosted inflow then smoothly joins the ring at the near side and spreads along the ring over the depletion time, returning to distributed star formation again.

To quantify the degree of the lopsided star formation, we divide the ring into two parts by a straight line $y = \tan(\phi) x$, where $\phi$ is the position angle of the dividing line measured counterclockwise from the positive $x$-axis.
We calculate the SFR separately in each part using the star particles with age younger than $10\,{\rm Myr}$, and then average it over 21 snapshots taken from $161$ to $181\,{\rm Myr}$ at $1\,{\rm Myr}$ interval.
We define the asymmetry parameter ${\cal A} (\phi)\,(\ge 1)$ as the (higher-to-lower) ratio of the averaged SFRs from the two parts for a given $\phi$.
We then repeat the calculations by varying $\phi$ to find the maximum value ${\cal A}_{\rm max} = \max_\phi {\cal A} (\phi)$.
Note that the position angle $\phi_0$ corresponding to ${\cal A}_{\rm max}$ differs for all models.

\begin{deluxetable}{lcc}
    \tablecaption{SFR asymmetry\label{tb:asym}}
    \tablehead{
\colhead{Model} &
\colhead{${\cal A}_{\rm max}$} &
\colhead{$\phi_0\,({\rm deg})$}
    }
    \startdata
{\tt constant} & 1.2 & $112$  \\
{\tt P15}      & 1.3 & $21$  \\
{\tt P50}      & 1.5 & $175$  \\
{\tt P100}     & 1.2 & $162$  \\\hline
{\tt asym}     & 1.8 & $167$  \\
{\tt off}      & 1.6 & $142$  \\
{\tt boost}    & 4.9 & $80$
    \enddata
\end{deluxetable}

Table \ref{tb:asym} lists the maximum asymmetry parameter ${\cal A}_{\rm max}$ and the corresponding position angle $\phi_0$.
Model {\tt constant} has ${\cal A}_{\rm max} = 1.2$ due to the randomness of star-forming positions under the symmetric inflows.
Models {\tt P15}, {\tt P50}, and {\tt P100} have similar or slightly higher ${\cal A}_{\rm max}$ than model {\tt constant} due probably to perturbations introduced by time variability of the inflow rate.
Models {\tt asym} and {\tt off} have ${\cal A}_{\rm max} =1.8$ and $1.6$, respectively, which are higher than the asymmetry parameter of model {\tt constant}  but still quite small considering a large asymmetry in the inflow rate.
In contrast, model {\tt boost} has ${\cal A}_{\rm max} = 4.9$, that is, the boosted mass inflow from the upper nozzle makes the SFR in the lower right side of the dividing line with $\phi_0=80^\circ$ higher by a factor of about 5 than in the opposite side, which is caused by the new star clusters in the boxed regions in Figure \ref{fig:asym}.
The similar asymmetry parameter for gas mass is almost unity for all models, because the newly accreted gas mass during one orbit $\dot{M}_\text{in}t_\text{orb}$ is smaller than the existing ring gas mass $\dot{M}_{\rm SF}t_\text{dep}\approx \dot{M}_\text{in}t_\text{dep}$, quickly spreading along the whole length of the ring within $t_\text{orb}$.
Our results suggest that asymmetric inflows alone are unable to  create lopsided star formation in the rings.
Rather, asymmetry in the ring star formation in our models requires a large (and sudden) boost in the inflow rate from one nozzle.

\section{Summary and Discussion}\label{s:discussion}

We perform semi-global numerical simulations of nuclear rings in which bar-driven mass inflows are represented by gas streams from two nozzles located at the domain boundaries.
To focus on what drives temporal and spatial variations of the ring SFR, we consider two series of models: one with time-varying inflow rate and the other in which the mass inflow rates from the two nozzles are differentially set.
Our simulations show both the mass inflow rate and SN feedback affect the ring SFR.
The oscillating inflow rate with period $\Delta \tau_\text{in}$ induces large-amplitude, quasi-periodic (with period equal to $\Delta \tau_\text{in}$) variations of the SFR, with the delay time of $\sim 10$--$20\,{\rm Myr}$, when $\Delta \tau_\text{in} \gtrsim 50\,{\rm Myr}$. 
During the delay time, gas accreted to the ring undergoes turbulent compression and/or gravitational contraction to increase its density above the threshold for star formation.

Unlike the mass inflow rate, the SN feedback is stochastic and responsible only for small-amplitude, short-term fluctuations of the SFR, with timescale $\lesssim 40\,{\rm Myr}$ (although  the SN rate also varies in response to a convolution of the SFR and stellar evolution delay time).
Since our standard definition of the SFR is proportional to the mass inflow rate averaged over a $10\rm\,Myr$ span, the effect of the inflow rate to the time variability of the SFR decreases with decreasing $\Delta \tau_\text{in}$.
Together with the stochastic effect of SN feedback, this makes the SFR almost independent of $\dot{M}_\text{in}$ for $\Delta \tau_\text{in}\lesssim 15\,{\rm Myr}$.
Asymmetry in the inflow rates from the two ends of a bar does not necessarily lead to asymmetric star formation in nuclear rings.
We find thar ring star formation is lopsided only a few Myr after the inflow rate from one nozzle is suddenly boosted by a large factor.
In what follows, we discuss our findings in comparison with observations.

\emph{Temporal Variation of the Ring SFR} -- 
The stellar age distributions in nuclear rings inferred from optical absorption spectra indicate that the ring star formation is likely episodic, with approximately $100\,{\rm Myr}$ timescales, rather than continuous \citep{allard06,sarzi07,gadotti19}.
It is uncertain what causes the observed variability of the ring star formation, yet existing theories and numerical simulations suggest that it is perhaps due to either SN feedback combined with fluid instabilities \citep{loose82,krugel93,elmegreen94,kruijssen14,krumholz17,torrey17} or the mass inflow rate \citep{wyseo13,wyseo14,wyseo19} which is known to vary with time \citep{wyseo19,sormani19,armillotta19,tress20}.
Our numerical experiments in the present work show that while both the SN feedback and the inflow rate can affect the ring SFR, only the latter can induce significant variations of the recent SFR with amplitude $\gtrsim 5$ over timescale $\gtrsim 50\,{\rm Myr}$.
Since star formation is almost random and widely distributed along a ring in our simulations, the resulting feedback is local and stochastic, driving only modest (within a factor $\sim 2$) variations of the SFR (see also \citetalias{moon21}) on short timescales.
Only if the local star formation events were temporally correlated throughout the ring would the resulting feedback simultaneously quench star formation and make the ring quiescent as a whole.
Our results therefore suggest that the intermittent episodes of star formation separated by $\sim 100\,\text{Myr}$ observed in nuclear rings are likely driven by variations in the mass inflow rather than the SN feedback.

In our models, variations in the SFR lead to variations in the SN rate, and this is reflected in time-varying total midplane pressure, gas scale height, and ISM weight since SN feedback is the main source of thermal and turbulent energy. We show that the predictions of the self-regulated equilibrium theory are satisfied in our simulations, even allowing for slow temporal variations (i.e. on a timescale longer than the local vertical dynamical time).  We also show that the time-varying depletion time agrees with the quasi-equilibrium prediction provided that an appropriate averaging window is used that accounts for the delay between feedback and star formation.

\emph{Lopsided Star Formation in Nuclear Rings} -- It has long been known that the star formation in the CMZ is asymmetric, such that most star formation occurs in positive longitudes, notably at Sgr B1 and B2 complexes \citep{bally10}.
Similar asymmetry has also been noted for the nuclear ring in M83 \citep{harris01,callanan21}.
Our second series (models {\tt asym}, {\tt off}, and {\tt boost}) offers a possible explanation for lopsided star formation.
The results of these models show that asymmetric inflows alone do not lead to lopsided star formation, because the gas from the upper and lower nozzles tend to be mixed up within a few orbital times, making the distribution of star particles in model {\tt asym} indistinguishable from that in model {\tt constant}.
A sudden decrease of the inflow rate from one of the nozzles in model {\tt off} does not create notable asymmetry, either.
However, when the inflow rate from one of the nozzles suddenly increases by a large factor, as in model {\tt boost}, the boosted inflow follows ballistic orbits and triggers enhanced star formation at the far side of the ring where the orbits converge, making the distribution of young clusters lopsided for a few Myr.

In real galaxies, boosted inflows may originate from fluid instabilities.
For example, a global simulation of \citet{sormani18} for the CMZ asymmetry found that the combination of wiggle and thermal instabilities creates dense clumps randomly distributed in the dust lanes.
Whenever the clumps infall along the dust lanes to the CMZ, the inflow rate becomes suddenly asymmetric and boosted by a large factor, making the gas distribution and hence star formation lopsided in the CMZ.
\citet{dale19} performed hydrodynamic simulations of an isolated, turbulent molecular cloud plunging into the CMZ, which may also represent a dense clump produced by the wiggle and thermal instabilities of the dust lanes.
\citet{dale19} showed that the compressive Galactic tidal force, as manifested by the orbit convergence near the pericenter passage, enhances the SFR at the downstream, qualitatively similar to what happens in model {\tt boost}.

We note that while our models can explain lopsided star formation, they do not show any noticeable asymmetry in the gas distribution which is observed in the CMZ and nuclear ring of M83.
In the case of M83, the asymmetric gas distribution might be caused by a recent minor merger, as indicated by an offset 
between the photometric and kinematic nucleus \citep{sakamoto04,knapen10}.
Another possibility is that the mass inflow occurs in the form of massive clumps rather than smooth streams, which may be caused by fluid instabilities \citep{sormani18} as mentioned above. 
To study the effects of such clumpy inflows on the ring SFR and gas distribution, it is necessary to run simulations that resolve density inhomogeneity, shear, and turbulent velocities in the dust-lane inflows.

The CMZ is known to harbor several prominent molecular clouds which are likely progenitors of massive star clusters \citep[e.g.,][]{hatchfield20}. It has been proposed that such clouds are parts of two spiral arms \citep{sofue95,sawada04,ridley17},
on either a closed elliptical orbit \citep{molinari11} or an open ballistic stream \citep{kruijssen15}. Although different orbital models place the clouds at different distances along the line of sight, they generally agrees that all the CMZ clouds including Sgr A--C, the \emph{brick}, and the \emph{dust ridge} clouds are situated at the near side of the CMZ.\footnote{Although the two spiral arm model seems to place the Sgr B2 complex and brick at the far side, the absorption and proper motion data strongly suggest that they are at the near side. \citet{ridley17} reconciled this inconsistency by suggesting that those clouds are kinematically detached, jutting out of the near-side arm.}
If these clouds are the results of recently boosted inflows from the far-side dust lane, the inflow rate might have been much lower in the past than the current value estimated by \citet{sormani19}, leading to the low SFR observed today.
If this is really the case, the CMZ might be on the verge of starburst in the near future \citep[see, e.g.,][]{longmore14,lu19}.

\emph{Comparison to Other Simulations} --
In our models the SN feedback alone induces a factor of $\sim 2$ fluctuations of the SFR with timescale $\lesssim 40\,\text{Myr}$ (see also \citetalias{moon21}).
This appears consistent with the results of the global simulations of \citet{sormani20} who found that the star formation in their simulated CMZ varies within a factor of $\sim2$.
In contrast, the global simulations of \citet{torrey17} for late-type, non-barred galaxies found that the star formation in the central $100\,{\rm pc}$ region goes through burst/quench cycles with the SFR varying more than an order of magnitude (fluctuations are lower on $\sim$ kpc scales).
Also, the global simulations of \citet{armillotta19} for a Milky Way-like galaxy found that the ring SFR varies more than an order of magnitude, although the most dominant cycle with period of $\sim50\,\text{Myr}$, driven by SN feedback, has an amplitude of $\sim5$.

It is uncertain what makes the SFR fluctuations in \citet{torrey17} and \citet{armillotta19} larger than those in our models and \citet{sormani20}.
But, we conjecture that the \emph{effective} feedback strength in the former might have been stronger than that in the latter due to low resolution.
In \citet{torrey17} and \citet{armillotta19}, the mass resolution ($\sim 10^3\,M_\odot$) is probably not high enough to resolve the Sedov-Taylor stage for most SNe exploding inside dense gas.
In this case, the feedback is in the form of momentum, amounting to an imposed value $p_*\sim 3$--$5\times 10^5\,M_\odot\,{\rm km\,s^{-1}}$ per SN \citep[in these simulations an approximate treatment of early feedback is also applied, which may increase the momentum budget]{hopkins18}.

When SNe are instead resolved, the feedback is in the form of energy and $p_*$ is self-consistently determined by the interactions of SN remnants with their surroundings.
When SNe occur in rapid succession, simulations with a cloudy interstellar medium show that $p_*$ per SN may be reduced relative to the single-SN value \citep[see][]{kor17}.
Also, \citetalias{moon21} found that a significant fraction of the total SN energy is \emph{wasted} in the ambient hot medium outside the ring, yielding $p_* \sim 0.4$--$0.8\times 10^5\,M_\odot\,{\rm km\,s^{-1}}$.
In test simulations (not presented in the paper) similar to model {\tt constant} that implement feedback by injecting momentum instead of energy, we found that  when $p_*$ is large, the SFR fluctuates with larger amplitudes.
Also, at lower mass resolution (for a given ring mass), the feedback energy from a given collapsed region will be a larger fraction of the total gravitational binding energy of the ring, and can therefore more easily disperse the ring.  
Thus, the amplitude of SFR fluctuations in simulations may be sensitive to both the specific parameter choices adopted for implementing feedback, and the resolution of the simulation.

\emph{Limitation of the Models} -- Our prescription of star formation and feedback lacks several physical elements that might affect the star formation history.
First, our models do not include early feedback mechanisms such as stellar winds and radiation, which can halt accretion and growth of sink particles before first SN explosions at $t\sim 4\,{\rm Myr}$.
Over the lifetime of a cluster, the momentum injection from SNe far exceeds that from winds and radiation, but this early feedback limits the star formation efficiency in individual molecular clouds \citep{rogers13,rahner17,jgkim18,jgkim21}.

Second, our spatial resolution is insufficient to resolve internal substructure within self-gravitating regions.
With collapse of internal overdensities and the resulting early feedback, the lifetime star formation efficiency in self-gravitating  structures might be lower, which could increase the overall gas density in the ring.
A higher density ring could be more prone to large-scale instability and star formation bursts.
Thus, the combination of higher resolution and additional feedback could potentially produce larger feedback-driven fluctuations.

\acknowledgments
The authors are grateful to the referee for a helpful report. 
The work of S.M.\ was supported by an NRF (National Research Foundation of Korea) grant funded by the Korean government (NRF-2017H1A2A1043558-Fostering Core Leaders of the Future Basic Science Program/Global Ph.D. Fellowship Program). The work of W.-T.K.\ was supported by the grants of National Research Foundation of Korea (2019R1A2C1004857 and 2020R1A4A2002885). The work of C.-G.K.\ and E.C.O. was supported in part by the National Science Foundation (AARG award AST-1713949) and NASA (ATP grant No. NNX17AG26G). Computational resources for this project were provided by Princeton Research Computing, a consortium including PICSciE and OIT at Princeton University, and by the Supercomputing Center/Korea Institute of Science and Technology Information with supercomputing resources including technical support (KSC-2021-CRE-0025).

\bibliography{mybib}
\end{document}